\documentclass[smallextended]{svjour2}                    % onecolumn
\smartqed  % flush right qed marks, e.g. at end of proof
\usepackage{amsfonts}
\usepackage{amsmath}
\usepackage{amssymb}
\usepackage{mathtools}
\usepackage{alltt}
\usepackage{subfigure}
\usepackage{mathptmx}
\usepackage{fancyhdr}
\usepackage{amsbsy}
\usepackage{textcomp}
\usepackage{array}
\usepackage{fixmath}
\usepackage[english]{babel}
\usepackage{graphicx}
\usepackage{color}
\usepackage{wasysym}
\usepackage{epstopdf}

\begin{document}

\title{Combining rare events techniques: phase change in Si nanoparticles.%\thanks{}
%about the article that should go on the front page should be
%placed here. General acknowledgments should be placed at the end of the article.}
}
%\subtitle{Do you have a subtitle?\\ If so, write it here}

%\titlerunning{Short form of title}        % if too long for running head

\author{Simone Meloni \and
        Sergio Orlandini \and 
        Giovanni Ciccotti
}

%\authorrunning{Short form of author list} % if too long for running head

\institute{
           Simone Meloni \at
           School of Physics, Room 302 EMPS-UCD, University College Dublin, Belfield, Dublin 4, Ireland \\
           Consorzio Interuniversitario per le Applicazioni di Supercalcolo Per Universit\`a e Ricerca (CASPUR), Via dei Tizii 6, 00185 Roma, Italy \\
           {To whom the correspondence should be addressed: s.meloni@caspur.it} 
\and
Sergio Orlandini \at
              Dipartimento di Chimica, Universit\`a  ``Sapienza'', P.le A. Moro 5, 00185 Roma, Italy \\
              Consorzio Interuniversitario per le Applicazioni di Supercalcolo Per Universit\`a e Ricerca (CASPUR), Via dei Tizii 6, 00185 Roma, Italy
%              \email{fauthor@example.com}           %  \\
%             \emph{Present address:} of F. Author  %  if needed
           \and
           Giovanni Ciccotti \at
           School of Physics, Room 302/B EMPS-UCD, University College Dublin, Belfield, Dublin 4, Ireland \\
           Dipartimento di Fisica and CNISM, Universit\a ``La Sapienza'', Piazzale Aldo Moro 2, 00185 Rome, Italy           
}

%\date{Received: date / Accepted: date}
% The correct dates will be entered by the editor

\maketitle

\begin{abstract}
We introduce a combined Restrained MD/Parallel Tempering approach to study the difference in free energy as a function of a set of collective variables between two states in presence of unknown slow degrees of freedom.

We applied this method to study the relative stability of the amorphous vs crystalline nanoparticles of size ranging between $0.8$ and $1.8$~nm as a function of the temperature. We found that, at variance with bulk systems, at low $T$ small nanoparticles are amorphous and undergo an amorphous-to-crystalline phase transition at higher $T$. On the contrary, large nanoparticles recover the bulk-like behavior: crystalline at low $T$ and amorphous at high $T$.

% \subclass{MSC code1 \and MSC code2 \and more}
\end{abstract}

\section{Introduction}
\label{sec:Introduction}
Modern methods for rare events allow the search of metastable states in complex systems, such as systems in condensed phase. Examples of these methods are the Temperature Accelerated Molecular Dynamics (TAMD) \cite{tamd}, with the associated Temperature Accelerate Monte Carlo (TAMC) \cite{tamc} and the Methadynamics \cite{metadynamics1,metadynamics2}. These methods are based on the introduction of an extended system consisting of the physical degrees of freedom (positions and momenta of the atoms) and a set of extra dynamical variables ${\bf z} = \{z_i\}_{i = 1, m}$ representing possible realizations of a set of suitable collective variables $\{\theta_i({\bf x})\}_{i = 1, m}$ characterizing the states of the system. From now on, we will refer to the space of the ${\bf z}$ variables as the $z$-space. The $\bf z$ variables are coupled to the atoms by a suitable potential (see Sec.~\ref{sec:Theory-RestrainedMD}) that forces the latter to stay in a configuration compatible with the restraint $\{\theta_i({\bf x}) \sim z_i\}_{i}$. From these simulations we can compute the probability density function $P({\bf z})$ to be at a given point in the $z$-space. $P({\bf z})$ is connected to the Landau free energy by the relation $F({\bf z}) = -k_B T \log P({\bf z})$, where $k_B$ is the Boltzmann constant.
%In all these methods, the evolution of the atomic degrees of freedom is coupled with that of the additional degrees of freedom. 
In these methods, the sampling of the configuration space is accelerated by biasing the dynamics of the ${\bf z}$ variables: the latter are forced to visit less probable regions of the $z$-space and, as a results, the atoms will also visit the corresponding regions in the configuration space. In  TAMD and TAMC this biasing consists in setting the $z$-temperature to a value much higher than the physical temperature. This makes it possible for the system to sample regions at high free energy, $F({\bf z})$. In Metadynamics, the biasing amounts to apply a history dependent biasing potential on the $z$ variables that forces the system to visit region of the $z$-space not visited before. In both cases, the ${\bf z}$ variables, and as a result the atoms, can more easily overcome free energy barriers present in the system.

\begin{figure*}
  \begin{center}
\includegraphics[width=0.95\textwidth]{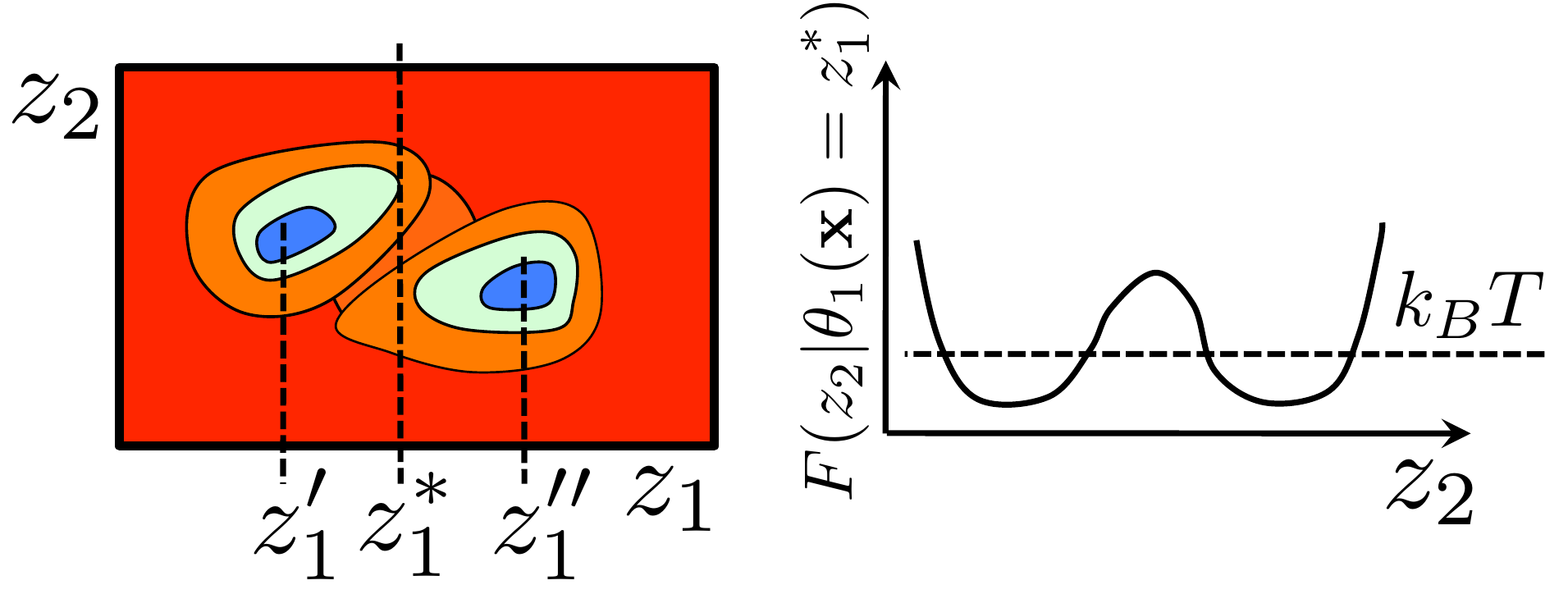}
 \end{center}
 \caption{(Left) Free energy landscape of a system with two slow degrees of freedom represented in colormap: blue $\rightarrow$ low free energy, red $\rightarrow$ high free energy. (Right) Free energy surface projected along $z_2$ at $\theta_1({\bf x}) = z_1^*$}\label{fig:2CV}
 \end{figure*}

Despite this acceleration, the reconstruction of the free energy surface $F({\bf z})$ by ``binning and histogramming'' is very expensive.
However, often we are interested only in the identification of metastable states and the characterization of their relative stability. This task can be accomplished by following a more efficient,  two steps approach. In the first step the $z$-space is scanned by relatively short TAMD/TAMC or Metadynamics runs. The corresponding $z$-space trajectory is analyzed  to find ``clusters'' \cite{clusterAnalysis,clusterAnalysis2}, i.e. those regions in the $z$-space where the trajectory variables tends to spend a significant amount of time. 
%(to this will correspond a maximum of the corresponding probability density function $P({\bf z})$).
These runs must be long enough to allow the identification of metastable states on $F({\bf z})$ but can be significantly shorter than the length required to accurately reconstruct the free energy surface.
In the second step, the difference in free energy of two of the metastable states identified in the previous step is calculated by numerically integrating the mean force $\nabla F({\bf z})$ along a path connecting them (thermodynamic integration). By repeating this operation for a proper set of pairs of states it is possible to reconstruct the relative free energy of all the metastable states. The $\nabla F({\bf z})$ can be estimated by computing the ensemble average conditional to $\{\theta_i({\bf x}) = z_i\}_i$ of a suitable observable (see Sec.~\ref{sec:Theory}). This task can be performed  by  restrained (constrained \cite{bluemoon}) MD or MC \cite{tamd,tamc}. This approach is based on the assumption that there are no free energy barriers in the domain orthogonal to the $z$-space, as otherwise the sampling of the conditional probability density function $\rho({\bf x}| \{\theta_i({\bf x}) = z_i\}_i)$ would be quite inefficient. We explain the problem of sampling $\rho({\bf x}| \{\theta_i({\bf x}) = z_i\}_i)$ by the following example. Consider a free energy surface of the type shown in Fig.~\ref{fig:2CV}. In this system there are two slow degrees of freedom, $\theta({\bf x})_1$ and $\theta({\bf x})_2$, but the two metastable states can be characterized by only $\theta({\bf x})_1$: the first corresponds to the condition $\theta({\bf x})_1 = z_1'$, and the second one corresponds to $\theta({\bf x})_1 = z_1''$. This means that, indeed, we can determine their relative stability by integrating $\nabla F(z_1)$ between $z_1'$ and $z_1''$. However, we notice that the free energy profile in the direction $z_2$ for a fixed value of $z_1$ ($F(z_2 | \theta_1({\bf x}) = z_1^*)$) presents a barrier higher than the thermal energy. As a result, the sampling of $\rho({\bf x} | \theta_1({\bf x}) = z_1^*)$ by restrained/constrained MD, or any other standard method, is biased.

In principle, this problem could be solved by including all the slow degrees of freedom in the $z$-space. However, already the identification of the collective variables characterizing the metastable states is very difficult and the identification of all the slow degrees of freedoms along the paths connecting these states often results just impossible.
In this paper we illustrate how it is possible, by combining restrained MD with the parallel tempering \cite{parallelTempering,parallelTempering2,MarinariParisi,PT-MD,parallelTempering3}, to compute the difference in free energy between two metastable states even in presence of unknown slow degrees of freedom. We apply this method to the study of the phase diagram of Si nanoparticles embedded in amorphous silica (a-SiO$_2$). We study the nature of the most stable phase as a function of the size and temperature of the nanoparticle. Anticipating our results, we found that, at variance with bulk Si, in small nanoparticles ($\diameter \leq 2.6$~nm) at low $T$ ($T \leq 1500$~K) the most stable phase is the disordered one. For larger nanoparticles the bulk-like behavior is recovered. 

The paper is organized as follows. In Sec.~\ref{sec:Theory} we introduce the combined restrained MD/Parallel Tempering method (RMD-TP). In Sec.~\ref{sec:ResultsAndDiscussion} we presents the results of the study of the phase diagram of the Si nanoparticles. Finally, in Sec.~\ref{sec:Conclusions} we draw some conclusions.

\section{Theoretical Background}
\label{sec:Theory}
This section is divided into two subsections. In the first one (Sec.~\ref{sec:Theory-RestrainedMD}) we revise the restrained MD method and demonstrate that, under the hypothesis that there are no slow degrees of freedom apart those listed in the set of collective variables $\{\theta_i({\bf x})\}_i$, this method allows to accurately and efficiently estimate the mean force $\nabla F({\bf z})$. In the second one (Sec.~\ref{sec:Theory-ParallelTempering}) we quickly summarize the parallel tempering method and explain how, by combining it with restrained MD, it is possible to efficiently compute the mean force $\nabla F({\bf z})$ even in presence of unknown slow degrees of freedom.

\subsection{Restrained MD}
\label{sec:Theory-RestrainedMD}
The restrained MD method considered in this paper is a specialization to the case of fixed $z$ of the TAMD method of Maragliano and Vanden-Eijnden \cite{tamd}. Let us introduce the following equation of motion for the atoms:
\begin{eqnarray}
\label{eq:TAMD}
m {\ddot {\bf x}} &=& -\nabla_x U^k({\bf x}, {\bf z}) + thermo(\beta)
\end{eqnarray} 

\noindent
where $m$ is the physical mass, $\beta = \left (k_B T\right )^{-1}$ and $thermo(\beta)$ indicates that the atoms are coupled to a thermostat at $\beta$. The potential $U^k({\bf x}, {\bf z}) = V({\bf x})  + \sum_{i=1}^m k/2 (\theta_i({\bf x}) - z_i)^2 $ is the sum of the physical and the restraining potential. Let us compute the following average:

%\begin{widetext}
%\begin{eqnarray}
\begin{eqnarray}
\label{eq:effectiveForce}
f^k_i({\bf z}) &=& \lim_{\tau \rightarrow \infty}{1 \over \tau} \int_0^\tau dt\,\,\, k(\theta_i({\bf x}(t)) - z_i)
  =  { \int d{\bf x}\,\,\, k(\theta_i({\bf x}) - z_i) \exp[-\beta U^k({\bf x}, {\bf z})] \over {\mathcal Z}^k({\bf z}) }\nonumber
\end{eqnarray}
%\end{eqnarray}
%\end{widetext}

\noindent where ${\mathcal Z}^k({\bf z}) = \int d{\bf x} \exp[-\beta U^k({\bf x}, {\bf z})]$. 
%the second equality is valid under the hypothesis that the dynamics of Eq.~\ref{eq:TAMD} is ergodic. 
The second equality in Eq.~\ref{eq:effectiveForce} stems from the assumption that, apart for the ${\bf z}$, the system is ergodic. $f^k_i({\bf z})$ is the $i$-th component of the gradient of $F_k({\bf z}) = -\beta^{-1} \log[{\mathcal Z}_k({\bf z})/\mathcal{Z}]$, where ${\mathcal Z} = \int d{\bf x} \exp[-\beta V({\bf x})]$ is the canonical partition function of the real system. 
Since $\mathcal Z$ is $z$-independent its introduction does not affect our argument but it is necessary for the interpretation of $F_k({\bf z})$ as a free energy. Noting that $\lim_{k\rightarrow \infty }\exp[ -\beta {k \over 2} (\theta_i({\bf x}) - z_i)^2 ] / \sqrt{(2\pi / \beta k)} = \delta(\theta_i({\bf x}) - z_i)$, in the limit of $k \rightarrow \infty$ $F_k({\bf z}) \rightarrow -\beta^{-1} \log [P({\bf z})] = F({\bf z})$.
%, where $P({\bf z}) =  \int d{\bf x} \exp[-\beta V^k({\bf x})] \prod_i \delta(\theta_i({\bf x}) - z_i) / \int d{\bf x} \exp[-\beta V^k({\bf x})]$ is the probability density function that $\{\theta_i({\bf x}) = z_i\}_i$ . 
In conclusion, the time average of Eq.~\ref{eq:effectiveForce} along the restrained MD of Eq.~\ref{eq:TAMD} is an estimate of $\nabla F({\bf z})$, i.e. the gradient of the free energy at the point $\bf z$.

\begin{figure*}
  \begin{center}
\includegraphics[width=0.95\textwidth]{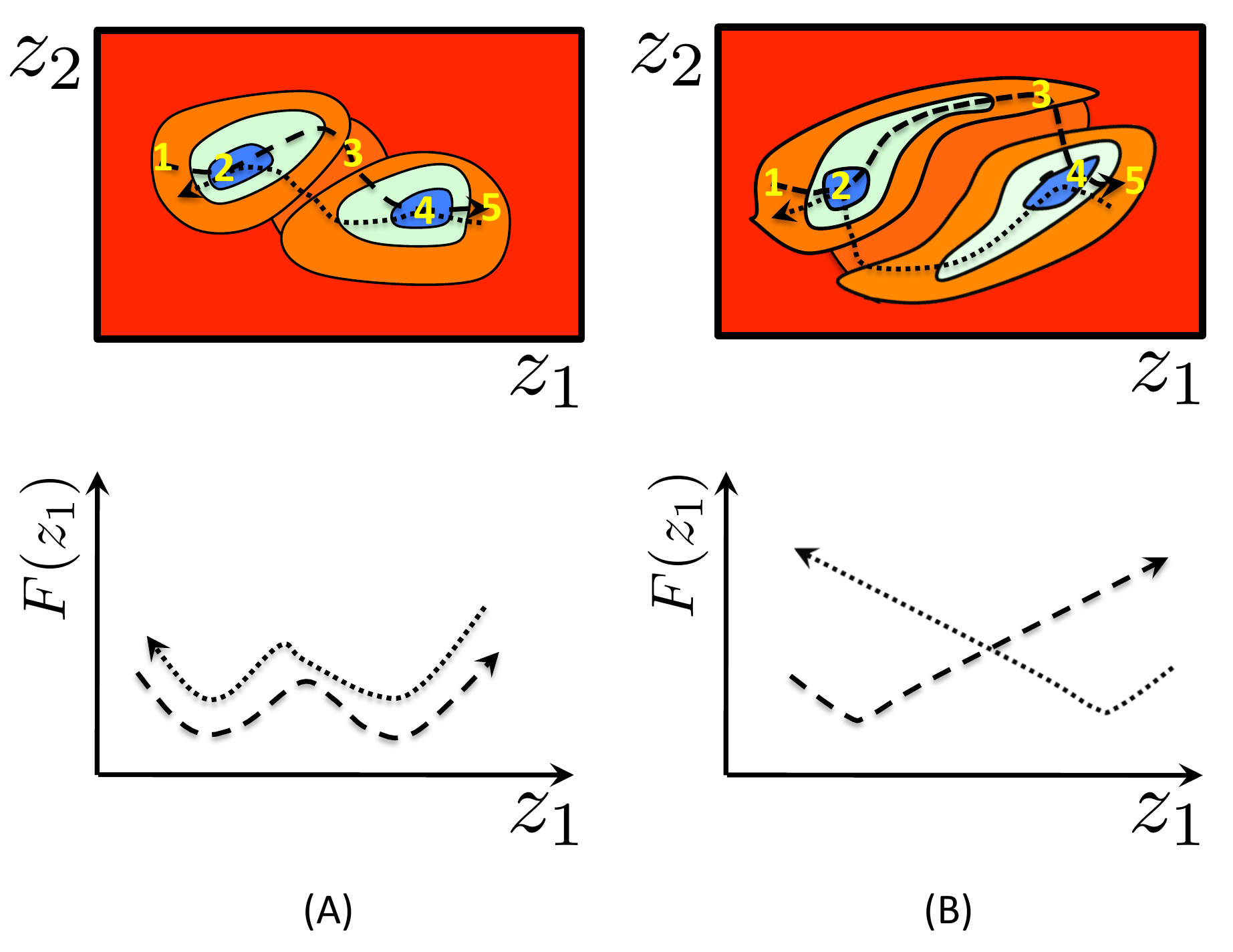}
 \end{center}
 \caption{(Top panels) Two cases of free energy landscape of a system with two slow degrees of freedom represented in colormap: blue $\rightarrow$ low free energy, red $\rightarrow$ high free energy. The dashed (dotted) curve represents the part of the space sampled by a pure restrained MD simulation going in the direction low-to-high (high-to-low) $z_1$. (Bottom panels) Corresponding free energy curves $F(z_1)$ as obtained from thermodynamics integration of the mean force computed by restrain MD. \label{fig:hysteresis}}
 \end{figure*}

Let us now discuss the case in which not all the slow collective variables have been identified. For sake of clarity, consider the case in which there are two collective variables, $\theta_1({\bf x})$ and $\theta_2({\bf x})$, but only the first is considered in the restrained MD calculation. We further assume that, even if $\theta_1({\bf x})$ is not the only slow collective variable, it is a good order parameter for the process under investigation. In this case the free energy profile might look like those reported in the top panels of Fig.~\ref{fig:hysteresis}. Let us focus first on the one shown in Fig.~\ref{fig:hysteresis}/A and let us compute $F(z_1)$ by performing the thermodynamics integration on only the $\nabla F(z_1)$ along $z_1$ as computed by restrained MD. If we start the calculation of the $\nabla F(z_1)$ from the minimum at low $z_1$, because of the presence of a free energy barrier along $z_2$, we shall be sampling only along the valley indicated by the dashed path. We keep sampling along this valley up to the point in which the barrier separating the two valleys is lower than the thermal energy. At this point the system will collapse in the other basin and, by increasing $z_1$, it will go toward the second minimum. Pushing the system forward along $z_1$ it will start climbing the free energy surface from the new basin. In short, we would compute the $\nabla F(z_1)$ only by sampling $\rho({\bf x}| \theta_1({\bf x}) = z_1)$ around the dashed path sketched in Fig.~\ref{fig:hysteresis}/A. The $\nabla F(z_1)$ computed along the segment $1$-$2$ of the dashed path will be negative and, correspondingly, the free energy will be decreasing. In the segment $2$-$3$ the $\nabla F(z_1)$ will be positive and the free energy increasing. This give raise to the minimum at low $Z_1$. In the segment $3$-$4$ and $4$-$5$ $\nabla F(z_1)$ will be negative and positive, respectively, and the free energy will present a second minimum. The free energy profile associated to this path is illustrated in the bottom panel of Fig.~\ref{fig:hysteresis}/A and is represented by a dashed line with an arrow indicating the direction along which the thermodynamic integration is performed. Let us not perform the same ideal experiment going in the other direction: from high to low $z_1$. In this case the path, denoted by a dotted line, will follow the other valley. As a result the free energy barrier is shifted even if the position of the minima is correctly computed. However, the relative free energy of the two metastable states is wrong following both the low-to-high and the high-to-low paths as the $\nabla F(z_1)$ was obtained from a bad sampling of $\rho({\bf x}| \theta_1({\bf x}) = z_1)$. 
However, we must consider also the case, shown in Fig.~\ref{fig:hysteresis}/B, in which the two valleys are very long and the point in which the system collapse in the other basin (segment $3$-$4$) is very close, or beyond, the second minimum (see dashed and dotted path in panel B). In the first case the mean force will be negative but small along this segment, while in the second case it will be positive. Depending on the exact placement of this segment, the reconstructed free energy profile might present one deep and one shallow minimum or just one minimum, as in the example of Fig.~\ref{fig:hysteresis}/B. This second example illustrates that also the qualitative picture of the free energy surface can be falsified if not all the slow degrees of freedom are taken into account, despite the ones considered could be the good order parameters for the process considered. In the next section, we will illustrate how this problem can be solved by combining restrained MD with the Parallel Tempering.

\subsection{Parallel Tempering and combined Restrained Parallel Tempering MD}
\label{sec:Theory-ParallelTempering}
In parallel tempering the simulated system is composed of $M$ replicas of the original system. Each replica is in the canonical ensemble at a temperature $T_i$ ($i$ is the index of the replica). Since the replicas do not interact with each other, the probability density function that the system is in the configuration ${\bf x}_1$ in the replica $1$, in the configuration ${\bf x}_2$ in the replica $2$, etc. is:

\begin{eqnarray}
\label{eq:ParallelTemperingPDF}
w({\bf x}_1,\dots,{\bf x}_M) =  {\prod_{i=1}^M \exp[-\beta_i V({\bf x}_i)] \over {\mathcal Q} (T_1, \dots, T_M)}
\end{eqnarray}

\noindent where $V({\bf x})$ is the potential governing the replicas and $\beta_i = 1 / k_B T_i$. ${\mathcal Q} (T_1, \dots, T_M) = \prod_{i=1}^M \int d{\bf x}_i  \exp[-\beta_i V({\bf x}_i)] $ is the overall configuration integral.
To sample $w({\bf x}_1,\dots,{\bf x}_M) $ we can use a MC procedure with two types of moves: (i) the standard MC moves within each replica (the specific type of moves depend on the system: atomic, molecular, etc.) and (ii) a move in which the positions of the atoms of two replicas are swapped. The first type of moves is accepted/rejected according to the usual Metropolis criterion:
\begin{eqnarray}
\label{eq:Metropolis}
a({\bf x}_i \rightarrow {\bf x}'_i) = min\left (1, {w({\bf x}_1,\dots, {\bf x}'_i,\dots, {\bf x}_M,)\over w({\bf x}_1,\dots, {\bf x}_i,\dots, {\bf x}_M,)}\right )
= min \left (1, \exp[- \beta_i (V({\bf x}'_i) - V({\bf x}_i))] \right) \end{eqnarray}

The second type of moves, which is attempted from time to time during the simulation, is accepted according to the probability 
\begin{eqnarray}
\label{eq:ParallelTempering-Metropolis}
a({\bf x}_i \rightarrow {\bf x}_j, {\bf x}_j \rightarrow {\bf x}_i ) &=& min\left (1, {w({\bf x}_1,\dots, {\bf x}_j,\dots,  {\bf x}_i,\dots, {\bf x}_M,)\over w({\bf x}_1,\dots, {\bf x}_i,\dots,  {\bf x}_j,\dots, {\bf x}_M,)}\right ) \nonumber \\
&=& min \left (1, \exp[-(\beta_j - \beta_i) (V({\bf x}_i) - V({\bf x}_j))] \right) \end{eqnarray}

\noindent where $i$ and $j$ are the indexes of the replicas for which the swap is attempted. Let us assume that $T_j >> T_i$. Then, the replica $j$ has an higher probability to overcome (free) energy barriers during the standard MC part of the simulation. Therefore, it has an higher probability to visit the many metastable states possibly present in the system. The swapping move, then, allows the low $T$ replica to sample these states too. This explains why, when the system presents energy barriers, parallel tempering with $M$ replicas is more efficient than standard MC of length $M \times N_{PT}$, where $N_{PT}$ is the number of steps of the parallel tempering simulation.

Parallel tempering can also be implemented using MD. In this case, the standard MC part of the simulation is replaced by a constant temperature MD. As in the MC based parallel tempering, the MD simulations of the replicas are carried out in parallel for a given number of steps after which  a swapping is attempted. This move is accepted/rejected according to the same $a({\bf x}_i \rightarrow {\bf x}_j, {\bf x}_j \rightarrow {\bf x}_i )$ probability reported in Eq.~\ref{eq:ParallelTempering-Metropolis}. In the MD based parallel tempering, after an accepted swapping, the momenta of the atoms are scaled by a factor  $\sqrt{T_{new}/T_{old}}$. Finally, both for MD and MC, after a swapping it is necessary to wait for the relaxation of the system at the new temperature before using its {\itshape trajectory} to sample $w({\bf x}_1,\dots,{\bf x}_M)$. 

In this work, we want to sample the probability density function of the system conditional to $\{\theta_i({\bf x}) = z_i^*\}_{i=1,m}$ (see previous section). Consistently with the restrained MD method explained in Sec.~\ref{sec:Theory-RestrainedMD}, this can be obtained by replacing the potential $V({\bf x})$ in Eq.~\ref{eq:ParallelTempering-Metropolis} with the potential $U^k({\bf x}, {\bf z})$ introduced in the previous section. For an adequate selection of the number of replicas and their temperatures, the piece-wise RMD-PT trajectories will sample the conditional probability density function $\rho({\bf x}| \{\theta_i({\bf x}) = z_i\}_i)$ also in presence of additional slow degrees of freedom.

\section{Application of the RMD-PT approach to the study of the phase diagram of embedded silicon nanoparticles.}
\label{sec:ResultsAndDiscussion}
In this section we present the results of our RMD-PT simulations aimed at studying the phase diagram of a Si nanoparticle embedded in a-SiO$_2$. By integrating the mean force computed according to the RMD-PT method along a path connecting the ordered and disordered states of a Si nanoparticle of a given size and at a given temperature, we are able to identify which is the most stable phase in the given conditions. The implementation of this approach requires the introduction of two collective variables, one controlling the size of the nanoparticle and the other controlling its degree of order. These collective variables are introduced in the Secs.~\ref{sec:Theory-CollVarr-Size} and \ref{sec:Theory-CollVarr-Q6}. In Sec.~\ref{sec:Theory-Setup} we describe the setup of our simulations. Finally,  in Sec.~\ref{sec:ResultsAndDiscussion-discussion} we present our results, and analyze and discuss the effect of the parallel tempering.

\subsection{The ${\mathcal R}({\mathbf x})$ collective variable to monitor the size of the nanoparticle}
\label{sec:Theory-CollVarr-Size}
We introduce the notion of size of the nanoparticle by a collective variable denoted by the symbol ${\mathcal R}({\mathbf x})$. In a system in which the nanoparticle is made of atoms of type `A' (Si in this case) and the matrix is made, or contains, atoms of type `B' (O in this case) a possible definition of the collective variable ${\mathcal R}({\mathbf x})$ is the distance between the center $x_c$ of the nano-particle (a point kept fixed during the simulations) and the closest oxygen atom, i.e. ${\mathcal R}({\mathbf x})=\min_i \{|{\bf x}_c -  {\bf x}^O_i|\}$, where ${\bf x}^O_i$ is the coordinate of the $i$-th oxygen atom. The force acting on the atoms associated to this collective variable cannot be straightforwardly evaluated since ${\mathcal R}({\mathbf x})$ is a non-analytical function of ${\mathbf x}$ and, therefore, there is no way to proceed through the direct calculation of the $\nabla k/2 ({\mathcal R}({\mathbf x}) - {\mathcal R}^*)^2$ term of Eq.~\ref{eq:TAMD}. We replace the above definition of the collective variable by a smooth explicit approximation of it that, in a proper limit, converges to its exact definition. This smooth explicit approximation is obtained in two steps: (i) first we obtain an explicit expression of $\min_i \{|{\bf x}_c - {\bf x}^O_i|\}$% as a function of the positions $x^O_i$, 
and (ii) then we introduce a smooth approximation to this expression. The first step consists in recognizing the following identity:

\begin{eqnarray}
\label{eq:Ranalytical}
\min_i\{|{\mathbf x}_c -  {\mathbf x}^O_i|\} &\equiv& \sum_{i}  \Big [ |{\mathbf x}_c -  {\mathbf x}^O_i|  
 \prod_{j \neq i}^{N_i} \Theta(|{\mathbf x}_c -  {\mathbf x}^O_j| - |{\mathbf x}_c -  {\mathbf x}^O_i|)  \Big ]
\end{eqnarray}

\noindent where $\Theta(x)$ is the Heaviside step function

\begin{equation}
\Theta(x) = \left \{ 
\begin{array}{ll}
0 & x \leq 0 \\
1 & x > 0
\end{array}
\right .
\end{equation}

 Let $l$ be the $O$ atom closest to the center of the nano-particle, then $  \prod_{j \neq i}^{N_i} \Theta(|{\mathbf x}_c -  {\mathbf x}^O_i| - |{\mathbf x}_c -  {\mathbf x}^O_j|) = \delta_{il}$  , where $\delta_{il}$ is the Kronecker symbol. 
This implies that the result of the r.h.s. of Eq.~\ref{eq:Ranalytical} is $ |{\mathbf x}_c -  {\mathbf x}^O_l|$, i.e. the distance from ${\mathbf x}_c$ of the closest O atom. Eq.~\ref{eq:Ranalytical} is, therefore, the explicit expression of the collective variable ${\mathcal R}({\mathbf x})$.  A smooth approximation to ${\mathcal R}({\mathbf x})$ can be obtained by replacing the Heaviside step function by a sigmoid function. We use a sigmoid function expressed in terms of the Fermi function:   
 
\begin{equation}
S(t) = 1 - {1 \over \ 1 + \exp[\lambda t]}
\end{equation}

\noindent where $\lambda$ is the parameter controlling its smoothness. In our simulations $\lambda$ has been chosen such that the sigmoid function goes from $0.95$ to $0.05$ in one atomic layer ($\sim 2$~\AA). A consequence of this replacement is that the size of the nano-particle is now defined as a weighted average of the distance of one atomic layer of oxygen atoms from the centre of the nano-particle.

\subsection{The ${\mathcal Q}_6({\mathbf x})$ collective variable to monitor the phase of the nanoparticle}
\label{sec:Theory-CollVarr-Q6}
We compute the free energy of a Si nano-particle embedded in silica as a function of its degree of order, as measured by the bond-orientational order parameter (${\mathcal Q}_6({\mathbf x})$) introduced by Steinhardt et al. \cite{PhysRevB.28.784} for bulk systems. In this work we us a modified version of the original definition adapted to the case of confined systems, as described below in detail. 

In general, ${\mathcal Q}_6({\mathbf x})$ is defined as
% the properly normalized square modulus of the vector ${\mathcal Q}_{6m}(x)$, where m=-6, \dots, +6:

\begin{equation}
\label{eq:AppQlTot}
{\mathcal Q}_6({\bf x}) = \left ( {4 \pi \over 2 \times 6 + 1}  \sum_{m=-6}^6 |{\mathcal Q}_{6m}({\bf x})|^2 \right )^{1 \over 2}
\end{equation}

\noindent where ${\mathcal Q}_{6m}({\bf x})$ is the normalized and weighted sum of ${\it q}^i_{6m}({\bf x})$ (defined below) which, in bulk systems, reads

\begin{equation}
\label{eq:AppQlm}
{\mathcal Q}_{6m}({\bf x}) =  {\sum_{i=1}^{N} N_i {\it q}^i_{6m}({\bf x}) \over \sum_{i=1}^N N_i}
\end{equation}

\noindent where $N$ is the number of atoms in the system, $N_i$ is the number of nearest neighbors of the atom $i$ and $m= -6, \dots, 6$. In the case of confined systems we limit the sum over $i$ to just the atoms belonging to the nanoparticle. Consistently with our definition of the size of the nanoparticle, we identify these atoms as those at a distance lower than ${\mathcal R}^*$ from the center of the nanoparticle (${\mathcal R}^*$ is the restrain value for the collective variable ${\mathcal R}({\bf x})$ used in the RMD-PT simulation). According to this definition, the ${\mathcal Q}_6m({\mathbf x})$ of the nanoparticle is: 

\begin{equation}
\label{eq:AppQlmNano}
{\mathcal Q}_{6m}({\bf x}) = \frac{ \sum_{i=1}^N N_i {\it q}_{6m}^i({\bf x}) \Theta({\mathcal R}^* - |{\bf x}_i^{Si} - {\bf x}_c|)}{\sum_{i=1}^N N_i}
\end{equation}

\noindent As for the case of the collective variable ${\mathcal R}({\mathbf x})$, the Heaviside step function is replaced by a sigmoid function  $S(t) = 1 - {1 / (1 + \exp[\lambda t])}$. 
%In this way, the collective variable is only on atoms which belong to the nano-particle according to the definition $|x_i^{Si} - x_c| \leq {\mathcal R}^*$.

The ${\it q}^i_{6m}(x)$ function appearing in Eq.~\ref{eq:AppQlmNano} is defined according to the following expression:

\begin{equation}
{\it q}^i_{6m}({\bf x}) =  {\sum_{j=1}^{N_i} {\it Y}_{6m}({ \hat x_{ij}}) \over N_i}
\end{equation}

\noindent where $ {\it Y}_{6m}({ \hat x_{ij}})$ is the spherical harmonic function of degree $6$ and component $m$ computed on the solid angle $\hat x_{ij}$ formed by the distance vector ${\vec x}_{ij} = {\bf x}_i - {\bf x}_i$ and the reference system. The sum runs over the $N_i$ nearest neighbors of the atom $i$. The sum over the $m$ component in Eq.~\ref{eq:AppQlTot} makes the collective variable ${\mathcal Q}_6({\bf x})$ rotationally invariant, i.e. independent on the orientation of the reference system. 

When the system is an ideal crystal and the temperature is $0$~K, the environment of all the atoms is the same and, therefore, ${\mathcal Q}_{6}({\bf x})$ is maximum as there is not interference among the ${\it q}^i_{6m}({\bf x})$. {The value of the ${\mathcal Q}_{6}({\bf x})$ depends on the structure of the crystal. For bulk Si ${\mathcal Q}_{6}({\bf x}) = 0.63$}.
On the contrary, in a perfectly disordered system the orientation of bonds is random and, therefore, there is complete interference among the ${\it q}^i_{6m}({\bf x})$, and ${\mathcal Q}_{6}({\bf x})$ is zero. However, even when the system is at finite temperature and its size is finite, this order parameter is still able to distinguish between a disordered and a crystalline phase, being its fluctuation ($\sigma_{{\mathcal Q}_6} \sim 0.005$) much smaller than the difference $\Delta {\mathcal Q}_6$ between the two states (typically $\Delta {\mathcal Q}_6 \geq 0.1$, see Sec. \ref{sec:ResultsAndDiscussion-discussion}). 

%Before concluding this section it is worth to mention that the ${\mathcal Q}_{6}$ collective variable, or its modifications, has been already used to study crystallization by atomistic simulations \cite{ReintenWolde1996,Auer2001,PhysRevLett.94.235703} and experiments \cite{Gasser2001}.

%
%The biasing potential related to ${\mathcal Q}_6(x)$ gives rise to additional contributions to the interatomic forces that include the term $\nabla {\mathcal Q}_6(x)$. We have therefore the problem of computing the derivative of an Heaviside step function, which would produce an impulsive force. We solved this problem by replacing the Heaviside step function $H(x)$ with a sigmoid function $S(x)$, which makes the force no longer impulsive.  For similar reasons, each term in the calculation of ${\it q}_{6m}^i(x)$ is also multiplied by a sigmoid function so as to eliminate impulsive forces when atoms enter and exit from the list of nearest neighbors of a given atom belonging to the nano-particle. In both cases, we defined the sigmoid function in term of the Fermi function 

\subsection{Simulation setup} 
\label{sec:Theory-Setup}
In the present investigation, the restrained MD is governed by the superposition of the Billeter et al. environment-dependent force field \cite{PhysRevB.73.155329} and the restraining potential $k/2({\mathcal Q}_6({\mathbf x}) - {\mathcal Q}_6^*)^2$ $+ k'/2({\mathcal R}({\mathbf x}) - {\mathcal R}^*)^2$. $k$ and $k'$ are the parameters controlling the degree of biasing and must be large enough such that along the MD the values of ${\mathcal Q}_6({\mathbf x})$ and ${\mathcal R}({\mathbf x})$ oscillate about the target values ${\mathcal Q}_6^*$ and ${\mathcal R}^*$, respectively. As for the Billeter et al. potential, its reliability in modeling equilibrium and dynamical properties of Si nanoparticles embedded in silica, of the Si/a-SiO$_2$ interface and Si nanowires has been already well established \cite{PhysRevB.73.155329,fischer:012101,ippolito:153109,PhysRevB.81.014203,tuma:193106}.

The samples are prepared by thermally annealing a periodically-repeated amorphous silica system and embedding Si nanograins (extracted from a well equilibrates either amorphous or crystalline bulk). The amorphous silica sample is prepared through the quenching-from-the-melt procedure, that is by cooling down very slowly a high temperature SiO$_2$ melt. The total system contains from $\sim 6000$ to $\sim 12000$ particles, corresponding to a nano-particle radius varying in the range $1-2$~nm. The samples are first thermalized at $300$~K in order to release possible stress at the Si/silica interface. Typically, during such a thermalization step, the nanoparticles slightly shrink. After this initial step, we impose the restraint on the size of the nanoparticles and thermalize the samples at the various target temperatures. After this treatment the samples are ready for the restrained simulations described above. 

In order to verify possible artifacts due to finite-size effects, we repeated the calculation of the mean force at few selected values of ${\mathcal Q}_6^*$ and ${\mathcal R}^*$ on samples at an increasing size of the silica matrix. We did not observe any significant difference in them (the differences were within the statistical error). This demonstrates that there are no finite-size effects in our free energy calculations. 

As for the Parallel Tempering temperatures, we used the following seven values: $300$~K, $500$~K, $750$~K, $1000$~K, $1250$~K, $1500$~K, $1750$~K.

\section{Results and discussion}
\label{sec:ResultsAndDiscussion-discussion}
In order to compute the phase diagram of the embedded Si nanoparticle, the first point is to find minima in the free energy $F({\mathcal Q}_6 | {\mathcal R})$, where $F({\mathcal Q}_6 | {\mathcal R})$ denotes the free energy as a function of ${\mathcal Q}_6$ for a Si nanoparticle of size ${\mathcal R}$. In the present case, since we deal only with two collective variables, and since we know from experimental and theoretical facts what is the range of the values of the collective variables (see below), we can directly proceed to the reconstruction of the free energy in a given interval of both ${\mathcal Q}_6$ and ${\mathcal R}$.
We compute the mean force at a set of points in the range ${\mathcal Q}_6 \in [0, 0.65]$ and ${\mathcal R} \in [0.8 , 1.8]$~nm. 
The rationale for the upper limit of the ${\mathcal Q}_6$ range is that the value of the bond-orientational order parameter for an ideal Si crystal is ${\mathcal Q}_6 = 0.63$ and, since from experiments and previous calculations it is known that Si crystalline nanoparticles assume a structure with a (distorted) diamond-like core and a disordered periphery \cite{daldosso_role_2003,Wakayama1998124,jap-83_2228,PhysRevLett.93.226104}, we expect the ${\mathcal Q}_6$ of crystalline nanoparticles be lower than this limit. The samples created according to the protocol described in Sec.~\ref{sec:Theory-Setup} confirm that the ${\mathcal Q}_6$ of crystalline nanoparticles is lower  than this upper limit. However, after a preliminary scanning of $\nabla_{{\mathcal Q}_6 } F({\mathcal Q}_6 | {\mathcal R})$ that allowed to identify the region of ${\mathcal Q}_6 $ containing the minima in the above domain, we restricted the calculations to a smaller range: $[0.04, 0.285]$, $[0.07, 0.35]$ and $[0.03, 0.38]$ for the nanoparticles of radius $0.8$, $1.3$ and $1.8$~nm, respectively.
As for the size of nanoparticles, the same experiments mentioned above report unusual phase transitions (disorder-to-order with growing $T$) for nanoparticles in the range $[0.5, 2.0]$~nm. We decided therefore to study nanoparticles of three size in this range: $0.8$, $1.3$ and $1.8$~nm.

\begin{figure*}
  \begin{center}
\includegraphics[width=0.95\textwidth]{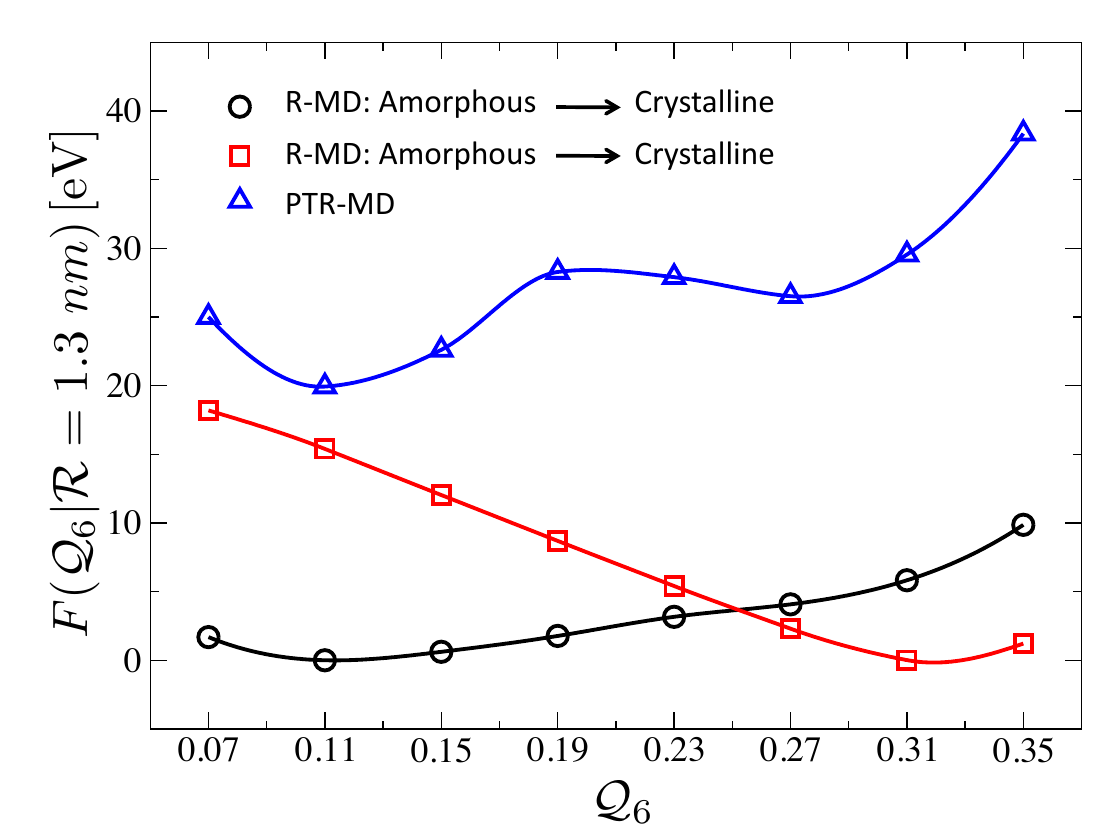}
 \end{center}
 \caption{Free energy vs ${\mathcal Q}_6$ curves for a nanoparticle of radius ${\mathcal R} = 1.3$~nm at $T = 750$~K. The free energy is calculated along the amorphous to crystalline ($\Circle$) and the crystalline to amorphous ($\Box$) paths, as described in the text. The free energy curve for the same nanoparticle size and system temperature as obtained from a RMD-PT simulation ($\triangle$) is also reported for comparison. An offset has been applied to this latter curves to improve readability. }\label{fig:FOfQ6-NoPTR}
 \end{figure*}

We initially performed our calculations without using the combined RMD-PT approach (i.e. purely restrained MD as described in Sec.~\ref{sec:Theory-RestrainedMD}). An example of the $F({\mathcal Q}_6 | {\mathcal R})$ function resulting from these calculation is shown in Fig.~\ref{fig:FOfQ6-NoPTR}, where the details of the simulation protocol are as follow. We start our simulations from one configuration either in the crystalline or amorphous domain (see Sec.~\ref{sec:Theory-Setup}), set the value of ${\mathcal Q}_6$ consistently with this configuration and, after a period of relaxation, compute the mean force at this value of ${\mathcal Q}_6$. The initial configuration for the next value of ${\mathcal Q}_6$ is selected randomly from the restrained MD run at the previous value of  ${\mathcal Q}_6$. Of course, a preliminary restrained MD is ran before computing the mean force, so as to let the system to relax to the new value of ${\mathcal Q}_6$. In the following, we will refer to the sequence of initial configurations as a ``path''. So, for example, the sequence selected starting from a low ${\mathcal Q}_6$ state and going to a high ${\mathcal Q}_6$ state will be referred to as $Amorphous \rightarrow Crystalline$ path.
The opposite path will be called $Crystalline \rightarrow Amorphous$. Back to the analysis of the purely restrained MD results shown in Fig.~\ref{fig:FOfQ6-NoPTR}, we notice that the mean force computed along the $Amorphous \rightarrow Crystalline$ path is different from that computed going in the backward direction. Moreover, the system presents only one minimum: in the amorphous domain when the mean force is computed along the $Amorphous \rightarrow Crystalline$ path, in the crystalline domain in the other case. 
%\textcolor{red}{Both results are in conflict with the observation that if we start a MD simulation in which we removed the restrain on ${\mathcal Q}_6$ from one configuration at low ${\mathcal Q}_6$ the system goes and stays in a disordered state, while if we run the same type of simulation staring from a configuration corresponding to a high ${\mathcal Q}_6$ the system evolves and remains in a crystal-like state. This observation indicates that the free energy as a function of ${\mathcal Q}_6$ is characterized by two wells.}
To tackle this situation, which clearly points out to an hysteresis of the second type illustrated in Sec.~\ref{sec:Theory-RestrainedMD}, we repeated the same simulation described above with the RMD-PT method and found that (i) the $ F({\mathcal Q}_6 | {\mathcal R})$ is independent of the path direction and  (ii) it presents two minima, one at low and one at high ${\mathcal Q}_6$ (see Fig.~\ref{fig:FOfQ6-NoPTR}).

\begin{figure*}
  \begin{center}
\includegraphics[width=0.95\textwidth]{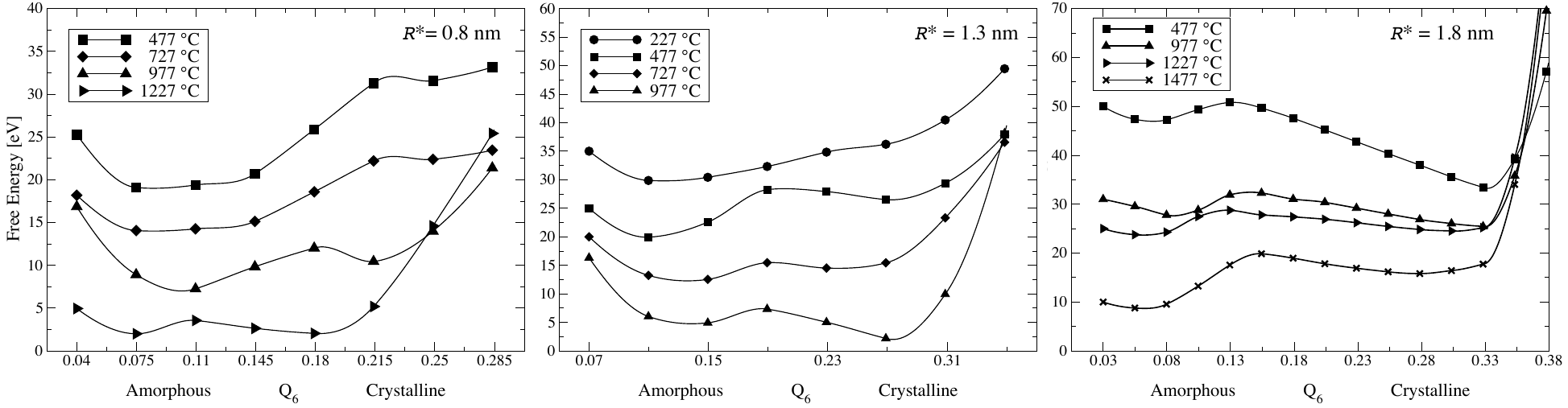}
 \end{center}
 \caption{Free energy vs ${\mathcal Q}_6$ curves for nano-particles with radius $0.8$~nm, $1.3$~nm and $1.8$~nm. The curves are shifted to improve readability.}\label{FOfQ6}
 \end{figure*}

\begin{figure*}
 \begin{center}
 \includegraphics[width=0.95\textwidth]{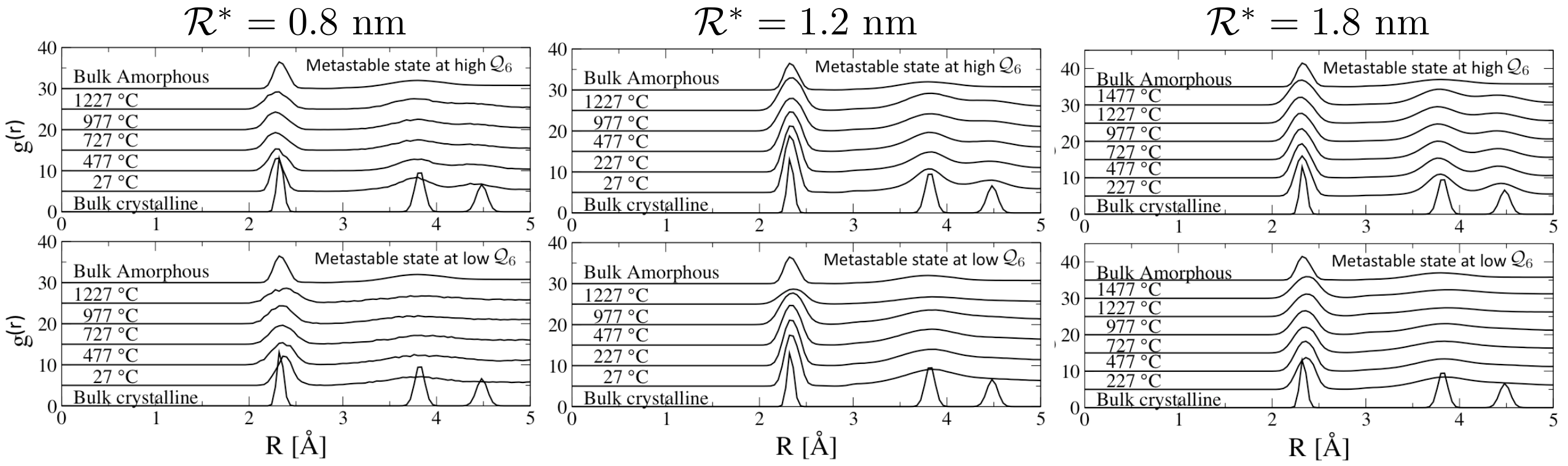}
 \end{center}
 \caption{Si-Si pair correlation function ($g(r)$) of the low and high ${\mathcal Q}_6$ metastable states at various $T$. Bulk crystalline and amorphous $g(r)$ are also reported for comparison.}\label{fig:GRVsT}
 \end{figure*}

We now move to the analysis of the results of our RMD-PT simulations. In Fig.~\ref{FOfQ6} the free energy curves $F({\mathcal Q}_6| {\mathcal R})$ of Si nano-particles of size ${\mathcal R} = 0.8$~nm, ${\mathcal R} = 1.3$~nm, and ${\mathcal R} = 1.8$~nm at various temperatures in the overall range $227 ^\circ$C - $1477 ^\circ$C are shown (the results are presented in Celsius for homogeneity with available experimental data). As a first remark, we notice that two metastable states are present at all temperatures and sizes. In general, to high values of  ${\mathcal Q}_6$ correspond crystalline states while to low values of the same order parameter correspond disordered (amorphous) states. However, especially for the smallest nanoparticle, the difference between the value of  ${\mathcal Q}_6$ corresponding to the two metastable states is small. Therefore, the identification of the phase corresponding to a given state requires a further investigation of the corresponding structure. We performed this analysis by computing the Si-Si partial pair correlation function $g(R)$ for the atoms belonging to the nanoparticle ($|{\mathbf x}^{Si}_i - {\mathbf x}_c| \leq {\mathcal R}^*$) averaging over configurations corresponding to the two metastable states. In the left-most panel of Fig.~\ref{fig:GRVsT} we report the $g(R)$ of the low (bottom panel) and high (top panel) ${\mathcal Q}_6$ metastable states for the ${\mathcal R} = 0.8$~nm nanoparticle at various temperatures. For the sake of comparison, we also show the $g(R)$ of bulk amorphous and crystalline states at $T = 627 ^\circ C$. %Let us start form the analysis of the structure of 
For the high ${\mathcal Q}_6$ metastable state, 
we notice that even at the highest $T$ the $g(R)$ is characterized by three peaks in the range $0 \leq R \leq 5$~\AA, one of which is sometimes barely visible. These peaks correspond to the bulk-like first, second, and third neighboring shell, respectively. 
They broaden and reduce in intensity by increasing the temperature; nevertheless, they are still well visible even at $T = 1227 ^\circ C$. In general, even at low $T$, these peaks are broader than the corresponding bulk crystalline ones. As for the low ${\mathcal Q}_6$ metastable state, we notice that in the same $R$ range analyzed for the high ${\mathcal Q}_6$ case there are only two peaks. The first one, sharp and intense, corresponds to the nearest neighbor Si-Si pairs. The second one, very broad and small, is usually assigned to the second neighbor pairs, which in amorphous system are distributed over a broad $R$ range. There is no other peak in the  $0 \leq R \leq 5$~\AA\ domain. Comparing the $g(R)$ of the low ${\mathcal Q}_6$ metastable state of this particle with amorphous bulk Si we notice that there is a one to one correspondence between the equivalent peaks in the two system. Similar results are found for the ${\mathcal R} = 1.3$~nm and ${\mathcal R} = 1.8$~nm nanoparticles (see the central and right-most panels of Figs.~\ref{fig:GRVsT}, respectively). On the basis of this analysis, we identified the high ${\mathcal Q}_6$ metastable state of all the nanoparticles at all temperatures as crystalline and the one at low ${\mathcal Q}_6$ as amorphous.

\begin{figure*}
 \includegraphics[width=0.95\textwidth]{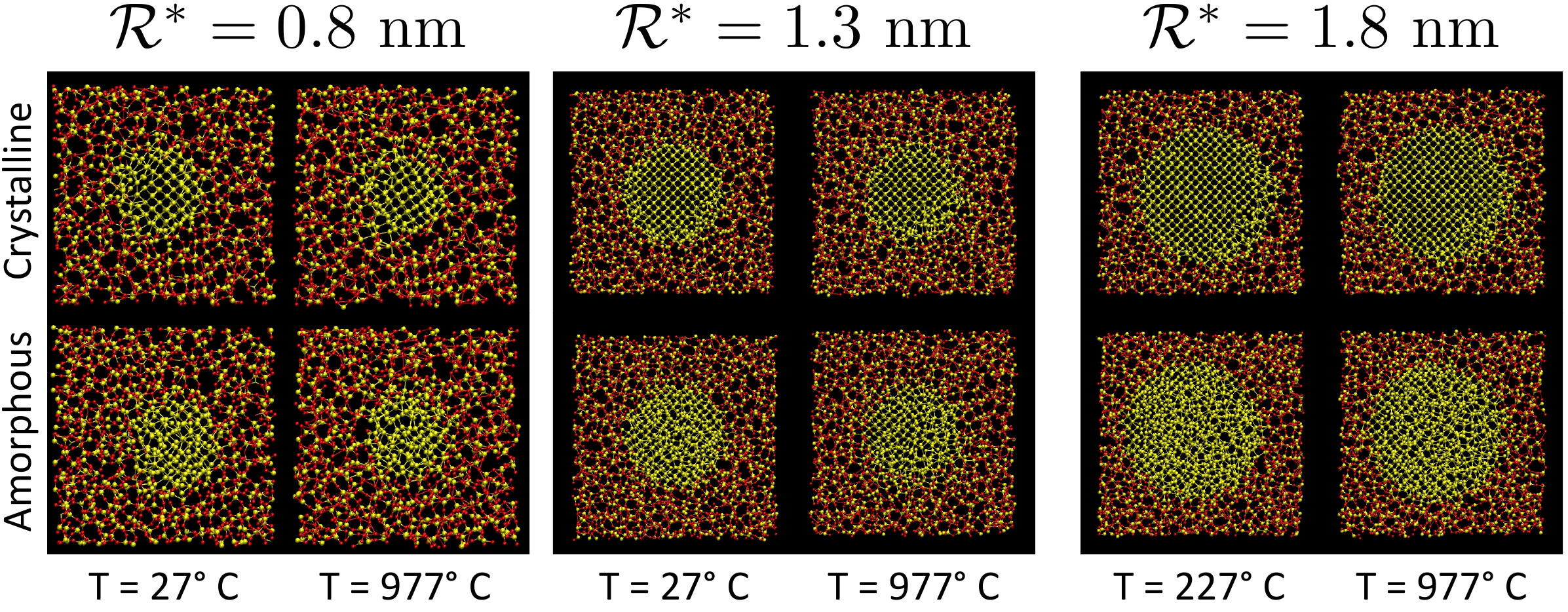}
 \caption{(color online) Snapshots of the nanoparticle of various size in the amorphous and crystalline metastable states at low and high temperature. Light (yellow) and dark (red) spheres represent the Si and O atoms, respectivelly.}\label{fig:snapshots}
 \end{figure*}

\textcolor{black}{The above conclusions are confirmed by the visual inspection of the structure of the nanoparticles in the low and high ${\mathcal Q}_6$ domains. In Fig.~\ref{fig:snapshots} we show few snapshots collected along our RMD-PT simulations at ${\mathcal Q}_6$ values corresponding to the minima of the free energy at low and high temperatures. It appears evident that the nanoparticles in the high ${\mathcal Q}_6$ domain have a crystal-like core in which the tetrahedral orientation of the Si-Si bonds is preserved. This core is surrounded by a disordered shell. This finding is consistent with previous results \cite{PhysRevLett.93.226104}. 
Instead, the structure of the nanoparticles corresponding to the low ${\mathcal Q}_6$ free energy minimum is completely disordered.}

% \begin{figure}
% \begin{center}
% \includegraphics[width=0.45\textwidth]{GRVsT-R=13.pdf}
% \end{center}
% \caption{Same as in Fig.~\ref{fig:GRVsT} for the nanoparticle of size ${\mathcal R} = 1.3$~nm.}\label{fig:GRVsT-R=1.3}
% \end{figure}
%
% \begin{figure}
% \begin{center}
% \includegraphics[width=0.45\textwidth]{GRVsT-R=18.pdf}
% \end{center}
% \caption{Same as in Fig.~\ref{fig:GRVsT} for the nanoparticle of size ${\mathcal R} = 1.8$~nm.}\label{fig:GRVsT-R=1.8}
% \end{figure}

\textcolor{black}{Having identified the nature of the metastable states we turn to the study of the phase diagram of Si nanoparticles with size and  temperature of the nanoparticle.} 
Our simulations (see Fig.~\ref{FOfQ6}) provide the following qualitative sharp picture: for small nano-particles (${\mathcal R} = 0.8-1.3$~nm) at low temperature (T $<  727 ^\circ$C) the most stable phase is the  amorphous one. The crystalline phase becomes equiprobable (${\mathcal R} = 0.8$~nm) or more probable (${\mathcal R} = 1.3$~nm) at higher temperatures. On the contrary, for larger particles (${\mathcal R} \geq 1.8$~nm) the behavior is bulk-like: at low temperatures (T $< 977 ^\circ$~C) the crystalline phase is the most stable one while at higher temperatures the amorphous phase is preferred . %For small nano-particles the equilibrium temperature (i.e. the temperature at which the free energy of the disordered and ordered phase are the same) decreases with the increase of the size of the nano-particle. This is indeed an effect of the steady increase of stability of the crystalline phase with respect to the disordered one with the size of the nano-particle.

%%%%%%%% 
Let us quickly summarize some experimental findings that we will use for comparison. Results from Energy Filtered Transmission Electron Microscopy (EFTEM) and Dark-Field Transmission Electron Microscopy (DFTEM) on Si-rich SiO$_x$ show that Si nanoparticles start to form at $1000 ^\circ$C \cite{jap-95_3723,jap-103_114303,Wakayama1998124,jap-83_2228,Wang2006486,jp-19_225003}. At this temperature they are all amorphous, while at $1100 ^\circ$C about one third become crystalline. By further increasing the temperature by $50 ^\circ$C, the fraction of crystalline nanoparticles rises up to ~60\%, while the average size of the nano-particles and the distribution of their size remains almost unchanged. Finally, at  $1250 ^\circ$C 100\% of nano-particles are crystalline. At this temperature the average size is slightly increased, but the particle size distribution still overlaps with the distributions at $1100 ^\circ$C and $1150 ^\circ$C.
%%%%%%
Our results provide the following interpretation of the these experimental results. At low temperatures the nanoparticles are small (as shown by experiments) and amorphous due to the found inversion of stability for nanoparticle of this size. At moderately higher temperatures, when size distribution of the nanoparticles is essentially unchanged, the larger nanoparticles in the sample transform from amorphous to crystalline, as the amorphous-to-crystalline transition temperature is lower for these particles. 
By further increasing the temperature the average size of the nano-particles increases significantly and almost all the nanoparticles transform into crystalline as at this size and temperature this is the most stable phase. The remaining small nanoparticles are crystalline too due to the inversion of stability. %This model is based on the experimental observation of the dependency of the average size and size distribution of the nanoparticles on the temperature \cite{jp-19_225003}. 
In other words, our results bring to the conclusion that the observed crystallization of the nanoparticles with the increase of the temperature is due to the interplay of the effect of the temperature on their size and the inversion of the relative stability of the amorphous and crystalline phase with the temperature for small nanoparticles.

\section{Conslusions}
\label{sec:Conclusions}

We have introduced a combined restrained MD/Parallel Tempering  approach to compute the difference in free energy as a function of a (a set of) collective variable between two metastable states in presence of unknown slow degrees of freedom.
We have applied the RMD-PT approach to the study of the phase diagram of Si nanoparticles embedded in amorphous silica as a function of their size and the system temperature. We found that, at variance with bulk Si, at low $T$ the small nanoparticles are amorphous, undergoing an amorphous-to-crystalline phase transition at higher $T$. Large nanoparticles instead recover, as expected, the bulk-like behavior; crystalline at low $T$ and amorphous at higher $T$. 
We also investigated the structure of the nanoparticles, finding that, in agreement with previous works \cite{PhysRevLett.93.226104}, the crystalline nanoparticles are formed by an ordered core surrounded by a disordered periphery.

\begin{acknowledgements}
The authors wish to acknowledge the SFI/HEA Irish Centre for High-End Computing (ICHEC) for the provision of computational facilities. G. C. and S. M. acknowledge SFI Grant 08-IN.1-I1869 for financial support. G. C. wish to also acknowledges financial support from Istituto Italiano di Tecnologia under the SEED project grant No. 259 SIMBEDD - Advanced Computational Methods for Biophysics, Drug Design and Energy Research.
S. M. acknowledges the support of the European Community under the Marie Curie Intra-European Fellowship for Career Development grant number 255406. Finally, S. O. acknowledges SimBioMa for financial support.
%If you'd like to thank anyone, place your comments here
%and remove the percent signs.
\end{acknowledgements}

% BibTeX users please use one of
%\bibliographystyle{spbasic}      % basic style, author-year citations
\bibliographystyle{spmpsci}      % mathematics and physical sciences
%\bibliographystyle{spphys}       % APS-like style for physics
%\bibliography{}   % name your BibTeX data base

% Non-BibTeX users please use

\end{document}